# NON-LINEAR PHENOMENA IN DUSTY PLASMAS

A. J. Turski, B. Atamaniuk and E. Turska
*Institute of Fundamental Technological Research*
*Polish Academy of Sciences, Warsaw, Poland*

Dusty plasmas are perhaps the fastest growing area in plasma physics with a surprisingly wide range of applications. They represent the most general form of space, laboratory and industrial plasmas.
We shall mainly discuss space dusty plasmas although many of the conclusions are valid for the laboratory plasmas as well. We use the term "dusty plasma" when the number of grains in Debye sphere is greater than one, and "dust in plasma" when the number of grains in Debye sphere is less than one.
The main objective of the paper is to determine asymptotic solutions to the initial-value conditions for Vlasov-Ampère/Gauss system of equations that is to find the "far field" solutions, [1, 2]. Next, we determine dispersion relations for longitudinal waves (DAW and DIAW) by use of the linearized Vlasov equations. In case of simplified equilibrium velocity distributions but for fully nonlinear plasmas, we determine velocity distributions $f_\alpha(u,\varsigma)$ where $\varsigma = x - Ut$ and by use of Sagdeev potential equations, we compute the solitary and double layer structures for a set of dusty plasma parameters. By use of Sagdeev potential, we determine the electrical capacity of plasma double layers.
It has been proved [3], that *Vlasov description of dusty plasmas is valid not only in the usual weakly coupled plasma regime but also in the strong-coupling limit* for dusty plasmas. Deviations from both limits are to be expected for the intermediate range of coupling when Coulomb crystallization occurs.

## 1. Introduction

Mathematical description of Dusty Plasmas is a very complex problem esp. of nonlinear processes. The main object is *the charge to mass ratio*. The ratio is a complicated function of plasma state and parameters, e.g. nonlinear waves, double layers, sheath, planetary rings, radial spokes, number densities of plasma components and other structures. There is no consequent and consistent description of Dusty Plasmas. Usually, we accept a model based on fixed charge/mass ratio or accept a probability distribution of the ratio and dust grains as an additional plasma parameter of dust component. It can be correct if the time scale of wave phenomena is much shorter than the time scale of charge/mass ratio changes.
In many instances the surface potential of a grain is approximately equal to $\Phi_s$=2.5 kT/e. However, the magnitude of the charge is not necessary equal to $4\pi\alpha|\Phi_s|$, where α is the grain radius. If the shape of a grain is irregular or the dust density becomes larger, the charge is smaller than this value. When secondary electrons are important, the surface potential can have three *equilibrium values*, and grains of different signs of charge can exist in plasmas. This may have important consequences for the rate at which grains collide and coagulate to form bigger particles. Such coagulation must have occurred during the early stages of the solar system evolution from its solar nebula stage.
In addition, the grain charge fluctuates randomly in a plasma and systematically as a grain gyrates about the magnetic field or moves through gradients of plasma density and/or temperature. These fluctuations can cause the angular momentum of a grain in a planetary magnetosphere to change and can lead to radial transport. It is evident that many more cases are relevant and need to be studied in detail.

## 2. Statement of the problem

We investigate the Vlasov-Ampere/Gauss system of equations for multispecies plasmas, that is

$$[\partial_t + u\partial_x + \frac{q_\alpha}{m_\alpha}E(x,t)\partial_u]f_\alpha(u,x,t) = 0 \quad \text{(Vlasov)} \tag{1}$$

$$\varepsilon_0 \partial_t E + \sum_\alpha q_\alpha \int_{-\infty}^{\infty} u f_\alpha du = 0 \quad \text{(Ampere)} \tag{2}$$

$$\varepsilon_0 \partial_x E + \sum_\alpha q_\alpha \int_{-\infty}^{\infty} f_\alpha du = 0, \ E = -\partial_x \phi \quad \text{(Gauss)} \tag{3}$$

let us assume

$$f_\alpha(u,x,t) = N_0^\alpha f_{0\alpha}(u) + \sum f_{n\alpha}(u,x,t) \quad \text{and} \ f_n \in O(E^n) \tag{4}$$

we derive hierarchy equations

$$(\partial_t + u\partial_x)f_{1\alpha} = -\frac{N_0^\alpha q_\alpha}{m_\alpha} E\partial_u f_{0\alpha}$$

$$\ldots\ldots\ldots\ldots\ldots\ldots\ldots\ldots\ldots\ldots\ldots\ldots\ldots\ldots\ldots\ldots \tag{5}$$

$$(\partial_t + u\partial_x)f_{n\alpha} = -\frac{q_\alpha}{m_\alpha} E\partial_u f_{n-1,\alpha}$$

and we search solutions for a given initial-value problem of the linear set of the hierarchy equations. The initial-value problem is

$$f_{1\alpha}(u,x,t_0) = g_\alpha(u,x), \ f_{n\alpha}(u,x=\pm\infty) = 0, \ E(x,t) = 0 \ for \ t \leq t_0, \tag{6}$$

and $f_{n\alpha}(u,x,t_0) = 0, \ for \ n = 2,3,...$

By use of Eqs (1) to (4) the Eq. (5) takes the form of the integro-differential abstract power series equation (Veinberg, Trenogin) [4]

$$E(x,t) + E_0(x,t) + P_g[G_\alpha(x,t)] + P[E(x,t)] = 0 \tag{7}$$

where $E_0(x,t)$ and $P_g[G_\alpha(x,t)]$ are linear and nonlinear terms, respectively. The terms are responsible for the initial disturbance $g_\alpha(u,x)$, see (6). $P[E(x,t)]$ is the nonlinear plasma response which does depend on the approved equilibrium distribution $f_{0\alpha}(u)$ and self-consistent field but it does not depend of the disturbance $g_\alpha(u,x)$. The crucial point is a convergence of the series (4) and of the integro-differential abstract power series terms $P_g[G_\alpha(x,t)]$ and $P[E(x,t)]$. We only note, that the problem is related to nonlinear Landau damping and instabilities. The far field solution is to be determined as $t_0 \to -\infty$. In that case, the terms $E_0(x,t)$ and $P_g[G_\alpha(x,t)]$ disappear and Eq. (7) becomes

$$E(x,t) + P[E(x,t)] = 0 \ as \ t_0 \to -\infty$$

We note, that in that case the series convergence of (4) and $P[E(x,t)]$ become more complicated from mathematical point of view as we have to do with improper integrals. It comes out from the fact of nonlinear Landau instabilities developing with the passage of time. If the series are convergent then the solution for particle velocity distributions takes the form

$$f_\alpha(u,x,t) = N_0^\alpha f_{0\alpha}(u + W_\alpha(u,x,t)), \tag{8}$$

where $W_\alpha(u,x,t)$ satisfies the following equation

$$[\partial_t + u\partial_x + \frac{q_\alpha}{m_\alpha}E(x,t)\partial_u]W_\alpha(u,x,t) = -\frac{q_\alpha}{m_\alpha}E(x,t) \tag{9}$$

Eq. (9) is a linear equation for $W_\alpha$ as $E$ is a given function. The solution can be determined by usual method of characteristics.

The relation (8) exhibits an equilibrium distribution memory of Vlasov plasmas. If one assumes the Maxwellian equilibrium distribution for "hot electrons" and a proper equilibrium distribution for "cold ions" then the far field solution does not exist, (Landau instabilities). A particular solution of Eq. (9) is

$$W_\alpha(\xi, u) = (u - U)\left[\left(1 + \frac{2q_\alpha \phi(\xi)}{m_\alpha (u - U)^2}\right)^{1/2} - 1\right] \quad (10)$$

where $\xi = x - Ut$ and $E(\xi) = -\partial_\xi \phi(\xi)$.

Assuming the Dirac delta equilibrium for cold plasma species, that is $f_{0c} = \delta(u)$ a stationary "*far field*" solution evolves into the form
$f_c(\xi, u) = \delta(u + W_c(\xi, u))$.

The well known "*cold particle*" number density is calculated as

$$n_c(\xi) = N_{0c} \int_{-\infty}^{\infty} \delta(u + W_c(\xi, u)) du = \frac{N_{0c}}{\left[1 - \frac{2q_c \phi(\xi)}{m_c U^2}\right]^{1/2}}. \quad (11)$$

In the case of" "*hot particles*", we accept "*square*" equilibrium distribution

$$f_{0h}(u) = \frac{1}{2a_h}\left[H(u + a_h) - H(u - a_h)\right]$$

and the hot particle number density takes the following form

$$n_h(\xi) = N_{0h} \int f_h(u, \xi) du = N_{0h} \frac{a_h + U}{2a_h}\left[1 - \frac{2q_h \phi(\xi)}{m_h(a_h + U)^2}\right]^{1/2} + N_{0h} \frac{a_h - U}{2a_h}\left[1 - \frac{2q_h \phi(\xi)}{m_h(a_h - U)^2}\right]^{1/2} \quad (12)$$

assuming $U/a_h \ll 1$, we have

$$n_h(\xi) \approx N_{0h}\left[1 - \frac{2q_h \phi(\xi)}{m_h a_h^2}\right]^{1/2} \quad (13)$$

By use of $n_c(\xi)$ and $n_h(\xi)$, we can determine Sagdeev potential and then calculate dust-ion-sound solitary waves for fully nonlinear plasmas.

### 3. Dispersion relation for linear waves

Assuming sufficiently small disturbances of plasma equilibrium, the solution of the linearized Vlasov-Ampere equations takes the form,

$$E(x, t) = E_0(x, t) + \int_0^t dt_1 \int E(x - x_1, t - t_1) K(x_1, t_1) dx_1 , \text{ where } K(x, t) = -\sum_\alpha \omega_\alpha^2 f_{0\alpha}(\frac{x}{t}). \quad (14)$$

We consider dust-electron-ion plasmas possessing the following equilibrium distributions:
$f_{0d}(u) = \delta(u)$ (*dust*), $f_{0e}(u) = \left[H(u + a_e) - H(u - a_e)\right]$ (*hot electrons*)
and $f_{0i}(u) = \left[H(u + a_i) - H(u - a_i)\right]$ (*hot ions*). Taking Fourier and Laplace transform of Eq. (14) with respect to $x$ and $t$, respectively, we obtain the dispersion relation for longitudinal plasma waves $D(k, s) \equiv 1 - K^e(k, s) - K^i(k, s) - K^d(k, s) = 0$, which takes the form

$$\omega^2 = \frac{\omega_e^2}{1 - \frac{k^2 a_e^2}{\omega^2}} + \frac{\omega_i^2}{1 - \frac{k^2 a_i^2}{\omega^2}} + \omega_d^2 \text{ , where } s = -i\omega.$$

Assuming rather cold ions, that is $\frac{k^2 a_i^2}{\omega} \ll 1$ and hot electrons $\frac{k^2 a_e^2}{\omega} \gg 1$ one obtains the following dispersion relation for the dust-ion-acoustic waves (DIAW)

$$\omega^2 \approx \frac{k_2 \lambda_{De}^2 (\omega_{01}^2 + \omega_d^2)}{1 + k^2 \lambda_{De}^2} \text{ , where } \lambda_{De}^2 = \frac{a_e^2}{\omega_{0e}^2}.$$

If $k^2 \lambda_{De}^2 \ll 1$, (long wave approximation), and $\omega_d^2 \ll \omega_{0i}^2$, we have the dust-modified ion-acoustic speed $C_s^2 = \frac{\omega^2}{k^2} \approx \frac{N_{0i}^2}{N_{0e}^2} \upsilon_s^2$ where $\upsilon_s = a_e (\frac{m_e}{m_i})^{1/2}$ and $q_d = Z_d q_e$.

The relation is similar to the usual ion-sound wave spectrum for non-isothermal plasmas that is $m_i a_i^2 \ll m_e a_e^2$. However in dusty plasmas, we usually have $T_e \approx T_i$ and then

$$\omega^2 \approx \frac{\omega_d^2 k^2 \lambda_{De}^2}{1 + k^2 \lambda_{De}^2 + \frac{N_{0i}}{N_{0e}}} \approx \omega_d^2 k^2 \lambda_{De}^2 \frac{N_{0e}}{N_{0i}} \text{ since } \omega^2 \ll k^2 a_i^2 \ll k^2 a_e^2 \text{ and}$$

$\frac{\lambda_{De}^2}{\lambda_{Di}^2} = \frac{N_{0i}}{N_{0e}} \gg 1 + k^2 \lambda_{De}^2$. The frequency $\omega$ of the dust-acoustic wave is very low and the dust-acoustic speed is $C_d^2 \approx \omega_d^2 \lambda_{De}^2 \frac{N_{0e}}{N_{0i}} = a_e^2 \frac{\omega_d^2}{\omega_e^2} \frac{N_{0e}}{N_{0i}}$.

In the virtue of the conditions: $a_i \ll \frac{\omega}{k} \ll a_e$ for DIAW as well as in view of the conditions $\frac{\omega}{k} \ll a_i \ll a_e$ for DAW, the waves are subjected to insignificant electron and ion Landau damping. These waves should have some relevance to low-frequency noise in the F-ring of Saturn.

## 4. Calculations of Sagdeev potentials and solitons

Let us introduce the following dimensionless normalization:
$y(\xi) = \frac{q_e \phi(\xi)}{m_i a_i^2}$ where $\xi$ is normalized to Debye length, $\lambda_{Di} = \frac{a_i}{\omega_{0i}}$, of ion plasma. The Mach number is $M = \frac{U}{a_i} \left[\frac{m_d}{Z_d m_i}\right]^{1/2} = \frac{U}{c_s}$ where the dust ion sound speed is $c_s = a_i \left[\frac{Z_d m_i}{m_d}\right]^{1/2}$. We denote ion-electron temperature ratio $R = \frac{m_i a_i^2}{m_e a_e^2} = \frac{T_i}{T_e}$ and assume global charge neutrality;

$N_{0i} = Z_d N_{0d} + N_{0e}$ hence $S_d = S = Z_d N_{0d}/N_{0i}$, $S_e = S = \frac{Z_d N_{0d}}{N_{0i}}$ and $S_e = S - 1 = \frac{N_{0e}}{N_{0i}}$.

Note that $m_d \gg m_i \gg m_e$ and $0 \le S \le 1$.

By Eqs (11) to (13) and the normalizations, we derive expressions for dust, ion and electron number densities; $\rho_d, \rho_i$ and $\rho_e$. Using the Gauss equation (3), we have

$$\frac{\partial^2 y}{\partial \xi^2} + \frac{1}{\varepsilon_o}\sum_\alpha \rho_\alpha = 0 \quad \text{and} \quad \frac{dV}{dy} = \frac{1}{\varepsilon_o}\sum_\alpha \rho_\alpha, \quad \alpha = e, i, d, \tag{15}$$

where V(y) is a Sagdeev potential. The energy integral can be obtained by integration of Eq. (15) and it takes the form

$$\frac{1}{2}\left(\frac{\partial^2 y}{\partial \xi^2}\right) + V(y) = 0, \quad \text{where} \quad V(y, M, S, R) = V_d(y, M, S) + V_i(y) + V_e(y, R)$$

and $V_d(y, M, S) = M^2 S\left[1 - (1 + \frac{2y}{M^2})^{1/2}\right] > 0$ for y<0,

$$V_i(y) = \frac{1}{3}\left(1 - (1 - 2y)^{3/2}\right) < 0 \text{ for y<0}, \quad V_e(y, S, R) = \frac{1-S}{3R}\left(1 - (1 + 2yR)^{3/2}\right) < 0 \text{ for y<0}.$$

The inertial dust term for y<0 delivers restoring force and thermal ions as well as electrons delivers wave pressures. We can expect negative potential solitons (rarefective solitons=antisolitons). The case S=1 ($S_e = 0$) represents the plasma where all the electrons are attached to the dust grains to form the two component plasma. Whereas the case S=0 ($S_e = 1$) is an electron-ion plasma. Numerical calculations are performed and we exhibit selected results in the graphical form.

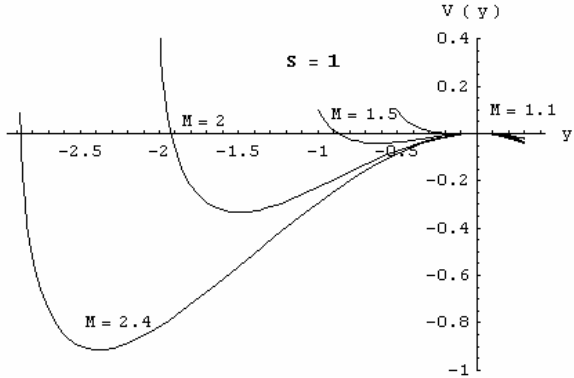

**Fig. 1** Sagdeev potentials V(y) versus y for Mach numbers M=1.1, 1.5, 2.0 and 2.4 as S=1.0 (no electrons). Soliton amplitudes y increase with increasing M.

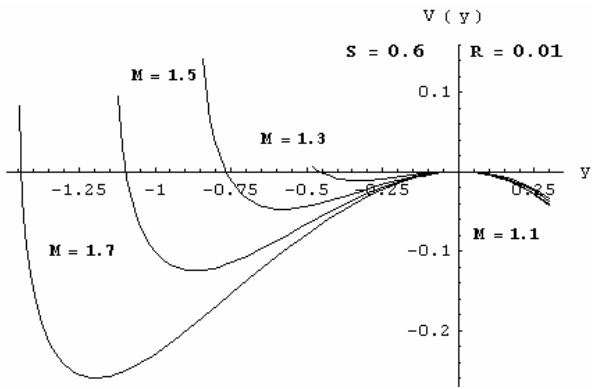

**Fig.2** Sagdeev potentials V(y) versus y for Mach numbers M=1.1, 1.3 and 1.7 as S=0.6. Non-isothermal plasmas; hot electrons and cold ions, (DIAW).

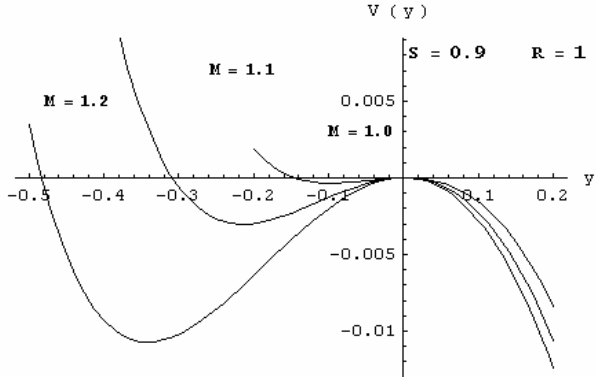

**Fig. 3** Sagdeev potentials V(y) versus y for Mach numbers M=1.0, 1.1 and 1.2 as S=0.1 and R=1. Isothermal plasmas, hot ions and electrons.

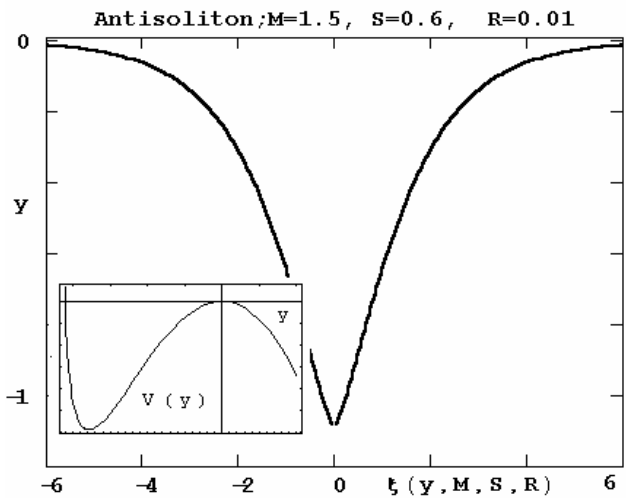

**Fig. 4** Sagdeev potentials V(y)-insert, and the respective soliton y($\xi$). Three component plasma; M=1.5, S=0.6 and R=0.01 (DIAW).

We conclude:
1. The initial value boundary problem is solved for "far field" solitons,(asymptotic solutions), in case of the fully nonlinear dusty plasma.
2. Landau damping that is wave-particle interactions have been excluded by choosing artificial equilibrium distributions.
3. The dispersion relation, here obtained, are the same like in the hydrodynamic description of dusty plasmas.
4. Due to the presence of negatively charged dust grains;
(i) Only negative potential solitons (rarefective solitons) can exist
(ii) Linear waves and solitons can exist in dust-ion- acoustic-wave plasmas (DIAW, R<<1, nonisothermal ions and electrons) as well as in dust-acoustic-wave plasmas (DAW, R $\approx$ 1, isothermal ions and electrons).
(iii) The existence of DAW solitons in the case of small amount of dust grains in plasmas (S=$S_d$=0.2) is secured for the soliton speeds less than the dust-ion-acoustic speed $c_s$, which was defined in this section.
5. Highly charged dust grains, ions and electrons considered. The presence of electron component ($S_e$ >0);
(i) lowers soliton amplitude as R is fixed,
(ii) lowers "lower limit" for M.

6. For small antisolitons ($|y| \ll 1$ and $S = 1$)) there is a full resemblance between hydrodynamic and kinetic antisolitons (K.d.V.). We note that $a_i \ll U \ll a_e$ and hence ion-acoustic dust solitons are subjected to insignificant electron and ion Landau damping.

## 5. Double layers

We assume four component plasmas to derive nonlinear structures called double layers. The following plasma components are accepted; cold dust grains, cold ions, hot ions and electrons and respective equilibrium velocity distributions are chosen. We obtain the Sagdeev potential composed of four components;
$$V(y) = V_d(y, M, S) + V_{ic}(y, M_c, S_c) + V_i(y, U_c) + V_e(y, R, A_e),$$
where $V_d, V_{ic}, V_i,$ and $V_e$ are the Sagdeev potentials for charged grains, cold ions, hot ions and electrons, respectively. We denoted M, S and R as in the section 4. New notations are;
$S_c = \dfrac{N_{0c}}{N_{0i}}$, $M_c = \dfrac{U_c}{a_i}$, and $U_d$, $U_i$, $U_c$, $A_e$ are the drift velocity of dust grains, hot ions, cold ions and electrons, respectively. The equations are analyzed numerically to determine the conditions leading to D.L. formation. The structure is subtle and very sensitive to parameter changes. We exhibit the Sagdeev potential for a double layer in dusty plasma.

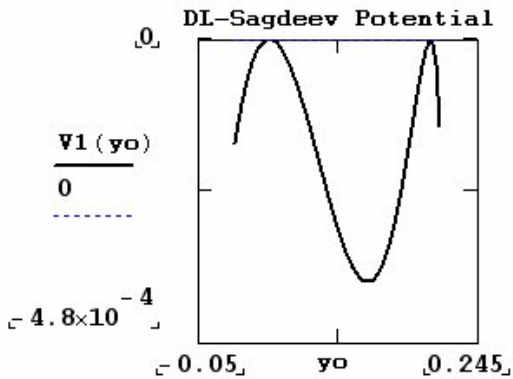

**Fig. 5** The Sagdeev potential versus dimensionless electric potential of 4-component dusty plasma. If $Z_d$=1000 and S=$Z_d$ $N_d/N_{01}$=1, $S_c/N_{0i}$, $M_c$ =$U_i/a_i \approx$ 1.05, R=0.2, $A_i$=1.3, $A_e$=1.3 then for the one dust drain, we have 1000 hot ions, 1350 cold ions and 1350 electrons (on an average)
.

## 6. Electrical capacity of D.L.

The charge for the unit plane surface of DL is determined as follow;
$$Q(y) = \int_{-\infty}^{\xi_p} \sum_\alpha \rho_\alpha(y(\xi)) d\xi = \varepsilon_0 \int_{-\infty}^{\xi_p} V'(y(\xi)) d\xi,$$ where $\xi_p$ is a point of inflexion of DL potential, that is a point of charge sign change. Integrating the Poisson equation, we derive the energy equation $\left(\dfrac{dy}{d\xi}\right)^2 + 2V(y) = 0$ hence $d\xi = \dfrac{dy}{\sqrt{-2V(y)}}$ and then the charge
$$Q(y) = \varepsilon_0 \int_{-\infty}^{\xi_p} V'(y(\xi)) d\xi = \varepsilon_0 \int_0^{y_p} \dfrac{V'(y)}{\sqrt{-2V(y)}} dy = \varepsilon_0 \sqrt{-2V_m(y_p)},$$ where "-$V_m(y_p)$" is the maximal magnitude of Sagdeev potential, which occurs for $y_p$. The DL capacity takes the form

$$C(y) = \frac{|Q|}{y_m} = \varepsilon_0 \frac{\sqrt{-V_m(y_p)}}{y_m},$$

where $y_m$ is a dimensionless potential jump of DL. The formula allows us to determine nonlinear dependence of DL capacity or charge as a function of potential jump. There are numerous applications of the relations. Let us mention the one, which is connected with plasma discharge due to DL occurrence at the electrode. The nonlinear capacity of DL leads to bifurcations and chaotic behavior of discharge currents.

**Acknowledgements** We are grateful to State Committee for Scientific Research ( KBN) for support through the grant No 2P03B-126-24.

## References

[1] A. J. Turski, B. Atamaniuk, K. Żuchowski, *„ Dusty plasma solitons In Vlasov plasmas"* , Arch. Mech. , **51**, 2, 167-179, 1999.
[2] A. J. Turski, B. Atamaniuk, *"Far field solutions of Vlasov-Maxwell equations and wave-particle interactions",* J. Tech. Phys., **2***,* 147-164, 1989.
[3] X. Wang, A. Bhattacharjee, *"On kinetic theory for strongly coupled dusty plasmas",* Phys. Plasmas, **3**, 4, 1189-1191, 1996.
[4] M.M. Veinberg, V.A. Trenogin, *"Theory of branching of solutions of nonlinear equations"*, Noordhoff Intern. Publishing, Amsterdam 1974.

E-mail addresses: aturski@ippt.gov.pl, batama@ippt.gov.pl and eturska@ippt.gov.pl